\documentclass[conference]{IEEEtran}

\newtheorem{definition}{Definition}
\newtheorem{lemma}{Lemma}
\usepackage[pdftex]{graphicx}
\usepackage{caption}
\usepackage{subcaption}
\ifCLASSINFOpdf
\else
\fi
\usepackage[cmex10, fleqn]{amsmath}
\usepackage{dsfont}
\DeclareMathOperator*{\argmax}{arg\,max}
\usepackage{algorithm,algorithmic}
\usepackage{url}

\usepackage{pgfplots}
\usetikzlibrary{patterns}

\begin{document}
%
\title{Mining Statistically Significant Attribute Associations in Attributed Graphs}

\author{\IEEEauthorblockN{Jihwan Lee}
\IEEEauthorblockA{Department of Computer Science\\
Purdue University\\
West Lafayette, IN\\
jihwan@purdue.edu}
\and
\IEEEauthorblockN{Keehwan Park}
\IEEEauthorblockA{Department of Computer Science\\
Purdue University\\
West Lafayette, IN\\
park451@purdue.edu}
\and
\IEEEauthorblockN{Sunil Prabhakar}
\IEEEauthorblockA{Department of Computer Science\\
Purdue University\\
West Lafayette, IN\\
sunil@purdue.edu}
}


\maketitle

\begin{abstract}
Recently, graphs have been widely used to represent many different kinds of real world data or observations such as social networks, protein-protein networks, road networks, and so on. In many cases, each node in a graph is associated with a set of its attributes and it is critical to not only consider the link structure of a graph but also use the attribute information to achieve more meaningful results in various graph mining tasks. Most previous works with attributed graphs take into account attribute relationships only between individually connected nodes. However, it should be greatly valuable to find out which sets of attributes are associated with each other and whether they are statistically significant or not. Mining such significant associations, we can uncover novel relationships among the sets of attributes in the graph. We propose an algorithm that can find those attribute associations efficiently and effectively, and show experimental results that confirm the high applicability of the proposed algorithm.
\end{abstract}


%
\IEEEpeerreviewmaketitle

\section{Introduction}
Nowadays graphs have emerged as a powerful abstract data type to represent and analyze complex data in a broad range of commercial and scientific applications including social networks~\cite{wasserman1994social, scott2012social}, bioinformatics~\cite{hu2002mining}, world wide web~\cite{broder2000graph, kleinberg1999web}, and so on. Mining structured patterns in graphs have been actively studied in the literature and such patterns including cliques~\cite{pei2005mining}, subgraphs~\cite{he2006graphrank, ranu2009graphsig, yan2008mining}, paths~\cite{scott2006efficient} and trees~\cite{chi2003indexing} help us better understand the intrinsic characteristics of graph data. Also, when the graph data come with auxiliary information such as node attributes, such information can be applied to various application areas, e.g., community detection, link prediction, graph clustering, network modeling, and etc. Thus, attributed graphs are more important than ever before to complex mining tasks.

While node attributes can be successfully employed to augment various mining tasks, the node attributes themselves could give us interesting patterns for better understanding graphs. Given an attributed graph where each node is associated with its attribute values, one might be interested in a pattern of node attribute values which co-occur between connected nodes. Let's call such co-occurred attribute values between two connected nodes an attribute association. This information can tell us directly the attribute patterns shared by connected nodes over the entire graph. In large scale, one might be interested in which attribute associations are most frequently observed or which attribute vector is most expected to be observed given another attribute vector in attribute associations. Looking at frequent attribute associations reveals the most dominant attribute associations in the graph by simply taking into account how many times they are held by connected nodes. Even though the frequent attribute associations give us which ones are dominant over the entire graph, they do not tell us which ones are really significant. That is because the frequency of an attribute association often does not depart from what we expect and therefore may not be meaningful actually if we already know the distributions of attribute values in the graph. Rather, identifying the statistically significant attribute associations where the pattern of the attribute association deviates from the expected, can potentially infer undiscovered possible relationships between nodes in the graph. The statistical significance of a pattern has been emphasized in various data mining problems~\cite{hamalainen2008efficient, gunnemann2012assessing, he2006graphrank, arora2014mining, sachan2012mining} and the previous works already explored why a statistically significant pattern is more important rather than a frequent pattern. Thus, in this paper we define a statistically significant attribute association and address the problem of uncovering it in attributed graphs.

\begin{figure}[t]
	\centering
	\includegraphics[width=0.4\textwidth]{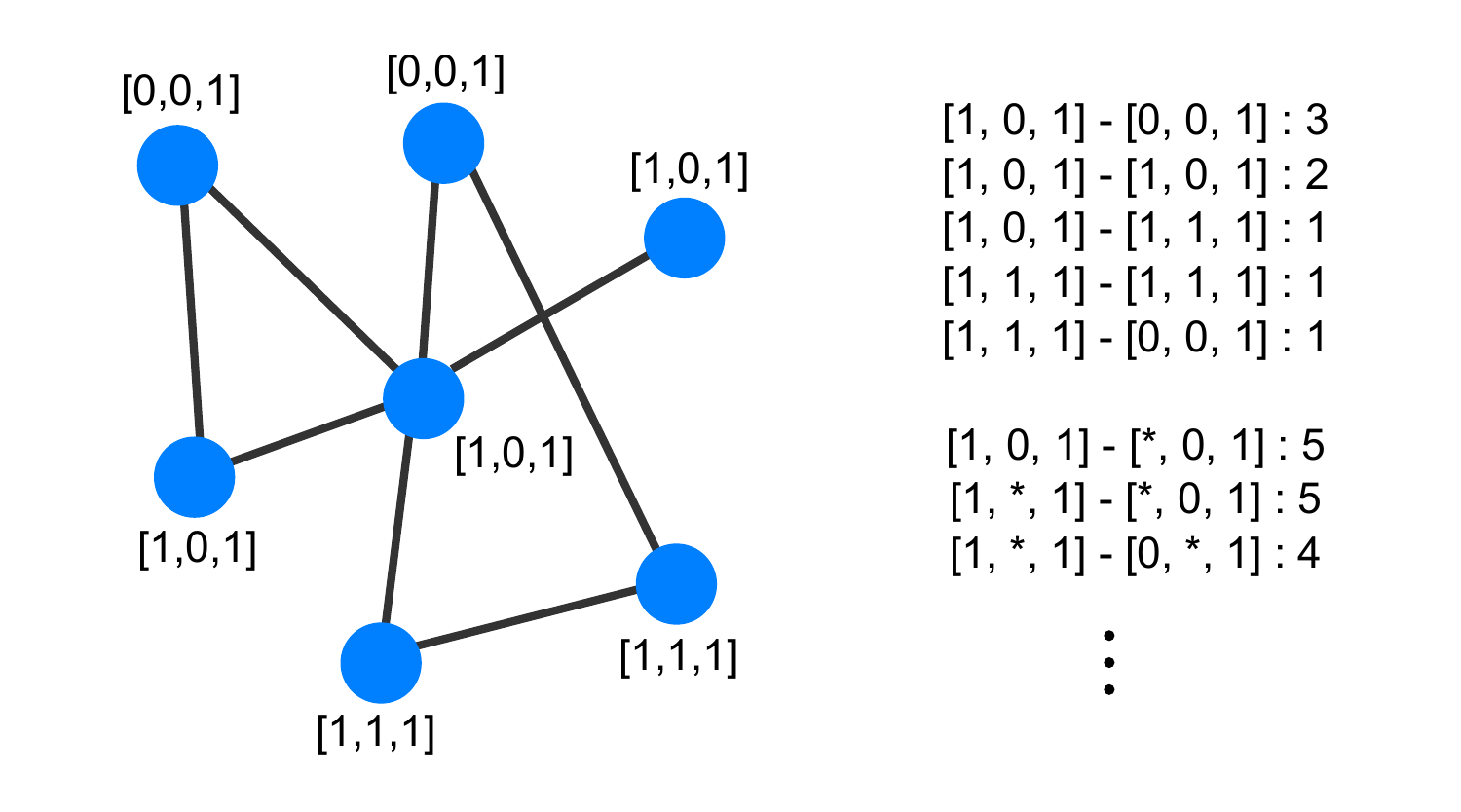}
	\caption{Attribute associations in attributed graph}
	\label{fig:intro_example}
\end{figure}

\figurename~\ref{fig:intro_example} shows an example that shows a list of possible attribute associations in an attribute graph. An attribute association is \textit{frequent} if the number of pairs of nodes is above a given threshold which is determined by \textit{freq\_support}. Unfortunately the frequency is not sufficient to measure the statistical significance of an attribute association since the frequency eventually depends on the actual distributions of the attribute values in the graph. We will closely see the set difference between the two in Section \ref{sec:experiments}. Also when obtaining significant associations, each attribute value does not always have to take discrete attribute value, e.g., 0 or 1 in binary case, as long as the association has enough statistical significance. Accordingly, we introduce wildcard attribute notation ($\ast$), which matches any value of the corresponding attribute.

The statistical significance of an attribute association with its frequency $k$ is determined by the probability that it is observed at least $k$ times or more, and the probability is called the \textit{p-value} of the attribute association. By measuring \textit{p-value}, we can identify the significant ones even though they are not frequent absolutely in the graph. Also, as shown in \figurename~\ref{fig:intro_example}, we are interested in even associations of partial attribute values as long as they are statistically significant. The main challenge of the problem is how to estimate the probability that an attribute association occurs in a random graph. There are as many different attribute associations as the number of edges in a graph, and if we consider even the partial attribute associations then the number of possible attribute associations grows exponentially. We address the challenge by transforming a graph $G$ into an alternative graph $\mathcal{AG}$, called \textit{association graph}, where each vertex contains a subset of nodes in $G$ that have the same or similar attribute values and each edge corresponds to a certain attribute association between two set of attribute values, each of which is represented by a cluster. During the process of transformation, we build $\mathcal{AG}$ such that the edges (i.e., associations) are statistically significant.

To experimentally evaluate our work, we use two real world attributed graphs. One is the \textit{DBLP co-authorship network} and the other is the \textit{Yelp social network}. We present the statistically significant attribute associations extracted from the graphs and compare them against the frequent attribute associations qualitatively. In addition to that, we show quantitatively how the statistically significant attribute associations can be used for boosting the performance of the link prediction task.

We summarize the contributions of our work as follows:

\begin{itemize}
	\item We formally define the novel problem of mining statistically significant attribute associations which aims to find patterns of co-occurred attribute values between nodes which deviate from the expected.
	\item We design and implement an algorithm that can find the statistically significant attribute associations efficiently and effectively.
	\item We conduct experiments using real world attributed graphs and show qualitative results as well as the actual application that can benefit from the results.
\end{itemize}

The paper is organized as follows. In Section~\ref{sec:related_work}, we introduce previous works related to our problem and discuss how our problem differs from them. In Section~\ref{sec:problem_statement}, we define the problem of mining statistically significant attribute association and provide basic background concepts. The novel algorithm to solve our problem is discussed in Section~\ref{sec:graph_transformation} and we present our experimental findings in Section~\ref{sec:experiments}. Finally, we conclude the paper in Sections~\ref{sec:conclusion}.

\section{Related Work}
\label{sec:related_work}
There are a number of previous works that have explored the statistical significance of patterns in various data mining and knowledge discovery tasks and have proposed efficient methods for mining the statistically significant patterns. \cite{arora2014mining, he2006graphrank} study the statistical significance of subgraphs where the nodes of the graph are labeled. \cite{arora2014mining} addresses the problem of finding statistically significant connected subgraphs in a vertex-labeled graph where the labels are discrete and continuous. The statistical significance is quantified by using the \textit{chi-square} statistic, which makes the na\"{\i}ve algorithm impractical because of the exponential number of subgraphs. They propose an efficient algorithm which converts the graph into a super-graph. In \cite{he2006graphrank}, the authors propose a technique for computing the statistical significance of frequent subgraphs in a graph database. In order to solve the difficulty of estimating the \textit{p-value} of a subgraph directly in the graph space due to the flexible structures of graphs, they tranform graphs into a feature space with predefined set of basis elements, and then approximate the significance of a feature vector in the feature space by using the binomial distribution. Although these two works explore the statistically significant patterns in graphs, they differ from our work in that they more focus on structured patterns, not attribute association patterns.

In addition to graphs, the statistical significance has been studied for other types of patterns as well. \cite{hamalainen2008efficient} extends the traditional association rule mining problem to searching statistically significant association rules such that some spurious rules are not included in the result set while considering statistical dependence. The significance of the observed frequency of an association rule is estimated by the binomial distribution. \cite{sachan2012mining} solves the problem of mining statistically significant substrings in a string generated from a memoryless Bernoulli distribution and uses the \text{chi-square} statistic as a quantitative measure of statistical significance. The statistical significance is considered for the sequential pattern mining problem as well in \cite{low2013mining}. The approach developed by the authors is able to efficiently mine unexpected patterns in sequence of itemsets without considering overlapping occurrences or conditioning the length of the sequence.
\section{Problem Statement}
\label{sec:problem_statement}
In this section, we give basic definitions of the attribute association, frequent association, statistically significant association, and define the problem of mining statistically significant attribute associations. Table~\ref{tab:notations} introduces the notations we use throughout the paper.

\begin{table}[!t]
	\renewcommand{\arraystretch}{1.3}
	\centering
	\begin{tabular}{| l | c |}
		\hline
		\bfseries Notation & \bfseries Meaning\\
		\hline
		$G=(V, E)$ & attributed graph \\ \hline
		$V = \{u_1, u_2, \dots, u_{|V|}\}$ & set of nodes in $G$\\ \hline
		$E$ & set of edges in $G$ \\ \hline
		$\mathcal{AG} = (\mathcal{V}, \mathcal{E})$ & association graph \\ \hline
		$\mathcal{V} = \{c_1, c_2, \dots, c_{|\mathcal{V}|}\}$ & set of clusters in $\mathcal{AG}$\\ \hline
		$\mathcal{E}$ & set of attribute associations in $\mathcal{AG}$ \\ \hline
		$\vec{a} = (a^1, a^2, \dots, a^l)$ & attribute vector of size $l$\\ \hline
		$\Delta$ & attribute association \\ \hline
		$\sigma$ & \textit{freq\_support} \\ \hline
		$\lambda$ & \textit{size\_support} \\ \hline
		$\delta_G$ & density of graph $G$ \\ \hline
		$\Psi_c$ & \textit{p-value} of cluster $c$ \\ \hline
		$TS(u, v)$ & tie-strength between node $u$ and $v$ \\ \hline
		$\Gamma(\cdot)$ & set of neighbors \\ \hline
		$\tilde{G_c}$ & subgraph of nodes within cluster $c$ \\ \hline
	\end{tabular}	
	\caption{Basic notations}
	\label{tab:notations}
	\vspace{-0.3cm}
\end{table}

\subsection{Attribute Associations}
\label{sec:attribute_association}
Suppose we have an attributed graph $G=(V, E, A)$ where $V=\{u_1, u_2, \dots, u_{|V|}\}$ is a set of nodes, $E = V \times V$ is a set of edges, and $A=\{\vec{a_{u_1}}, \vec{a_{u_2}}, \dots, \vec{a_{u_{|V|}}}\}$ is a set of attribute vectors, each of which is associated with a node in $V$. The attribute vector $\vec{a_u}$ of the node $u$ that holds $l$ different attributes is represented by a vector of $l$ binary values in that each binary indicates whether the node $u$ actually has a value for the corresponding attribute (in case of an $m$ multi-valued attribute, it can be transformed into $m-1$ dichotomous variables each with binary). Then we define an \textit{attribute association} between a pair of attribute vectors $\vec{a_1}$ and $\vec{a_2}$ as follows:

\begin{definition}
	\textit{Given two attribute vectors $\vec{a_1}=(a_1^1, a_1^2, \dots, a_1^l)$ and $\vec{a_2}=(a_2^1, a_2^2, \dots, a_2^l)$, the attribute association between them, denoted by $\Delta_{\vec{a_1},\vec{a_2}}$, is defined as a pair of two sets of attribute values, $\{i | a_1^i = 1\}$ and $\{i | a_2^i = 1\}$ where $i \in \{1, 2, \dots, l\}$.}
\end{definition}

Note that the attribute association is symmetric with respect to a given pair of attribute vectors $\vec{a_1}$ and $\vec{a_2}$, that is, $\Delta_{\vec{a_1},\vec{a_2}} = \Delta_{\vec{a_2},\vec{a_1}}$. Every pair of nodes has its attribute association and therefore there are as many attribute associations as the number of edges in $G$. The attribute association information is widely used in many different applications. For example, the link prediction algorithms that aim to predict whether a link will be newly formed between two unconnected nodes in the future usually employ the link structure information around the two nodes but it could leverage from using the attributes of the nodes as well. Many previous researches have shown that nodes in a graph tend to establish homophily or heterophily relationships in terms of their attributes~\cite{kim2012multiplicative, rogers1970homophily, mcpherson2001birds}. Another example of using attribute information is the community detection problem. Many early approaches to detect latent communities rely on only the link structure of a graph~\cite{coscia2012demon, yang2013overlapping, blondel2008fast}. That is, they detect communities such that nodes within the same community interact with each other more frequently than with those outside the community. However more recent studies use the node attributes as well as the link structure and show that the attribute information is helpful for community detection~\cite{balasubramanyan2011block, gunnemann2013efficient, ruan2013efficient}.

If an attribute association $\Delta$ is repeatedly observed and its frequency is over a given threshold $\sigma$ that is referred as \textit{freq\_support}, then we say $\Delta$ is a frequent attribute association.

\begin{definition}
	\textit{Given an attribute association $\Delta$ and a support $\sigma$, $\Delta$ is called a frequent attribute association if $fr(\Delta) \geq \sigma \times |E|$ where $fr(\Delta)$ is the number of pairs of nodes with $\Delta$.}
\end{definition}

When a frequent attribute association is given, we can say that there are many pairs of nodes having the association but it does not necessarily mean that the attribute association is really interesting. For example, in a social network of \textit{Purdue University Almuni}, it is not surprising to observe many connected nodes have the attribute association of \{``\texttt{Purdue}'', ``\texttt{CS}''\} -- \{``\texttt{Purdue}'', ``\texttt{CS}''\}. So we are interested in statistically significant attribute associations rather than frequent ones, which will be discussed in the following section.

\subsection{Statistically Significance}
\label{sec:statistically_significance}
The statistical significance of an object can be quantified by estimating the probability of the observed or rarer objects under the null hypothesis. Let $\delta_G$ denote the density of $G$ which is defined as the fraction of the number of edges in $G$ over all pairs of nodes ($\delta_G = \frac{|E|}{1/2 \cdot |V| \cdot (|V|-1)}$). If we randomly select two groups of nodes no matter which attribute values they have, denoted by $C_1$ and $C_2$ respectively, then the expected number of edges between $C_1$ and $C_2$ is $e(C_1, C_2) = |C_1| \cdot |C_2| \cdot \delta_G$ by assuming the probability of a pair of randomly selected nodes being connected to each other follows $\delta_G$. Also, assuming the edges are independent of each other, the actual number of edges $M$ between $C_1$ and $C_2$ would follow the binomial distribution with parameters $n = |C_1| \cdot |C_2|$ and $p = \delta_G$, and thus the probability of getting exactly $k$ edges among $n$ possible edges is given by the following probability mass function:

\begin{eqnarray}
	f(k; n, p) = P[M = k] = \binom{n}{k} p^k (1-p)^{n-k}
\end{eqnarray}

If each of $C_1$ and $C_2$ is a group of nodes with the same attribute values in $G$ which are specified by an attribute vector, then the attribute vectors $\vec{a_1}$ and $\vec{a_2}$ can be instantiated from $C_1$ and $C_2$ respectively and the attribute association between two attribute vectors is induced from the edges across the nodes of $C_1$ and the nodes of $C_2$. So we can measure the statistical significance of a given attribute association $\Delta_{\vec{a_1}, \vec{a_2}}$ based on the probability $P[M \geq k]$ that the observed or higher number of edges occur between $C_1$ and $C_2$ in which the nodes have $\vec{a_1}$ and $\vec{a_2}$ respectively. The association is said to be statistically significant if the estimated probability $P[M \geq k]$ is very small.

\begin{eqnarray}
	\label{eq:pval_association}
	P[M \geq k] = 1 - \displaystyle \sum_{i = 0}^{k-1} \binom{n}{i} p^i (1-p)^{n-i}
\end{eqnarray}

\begin{figure}[t]
	\centering
	\includegraphics[width=0.45\textwidth]{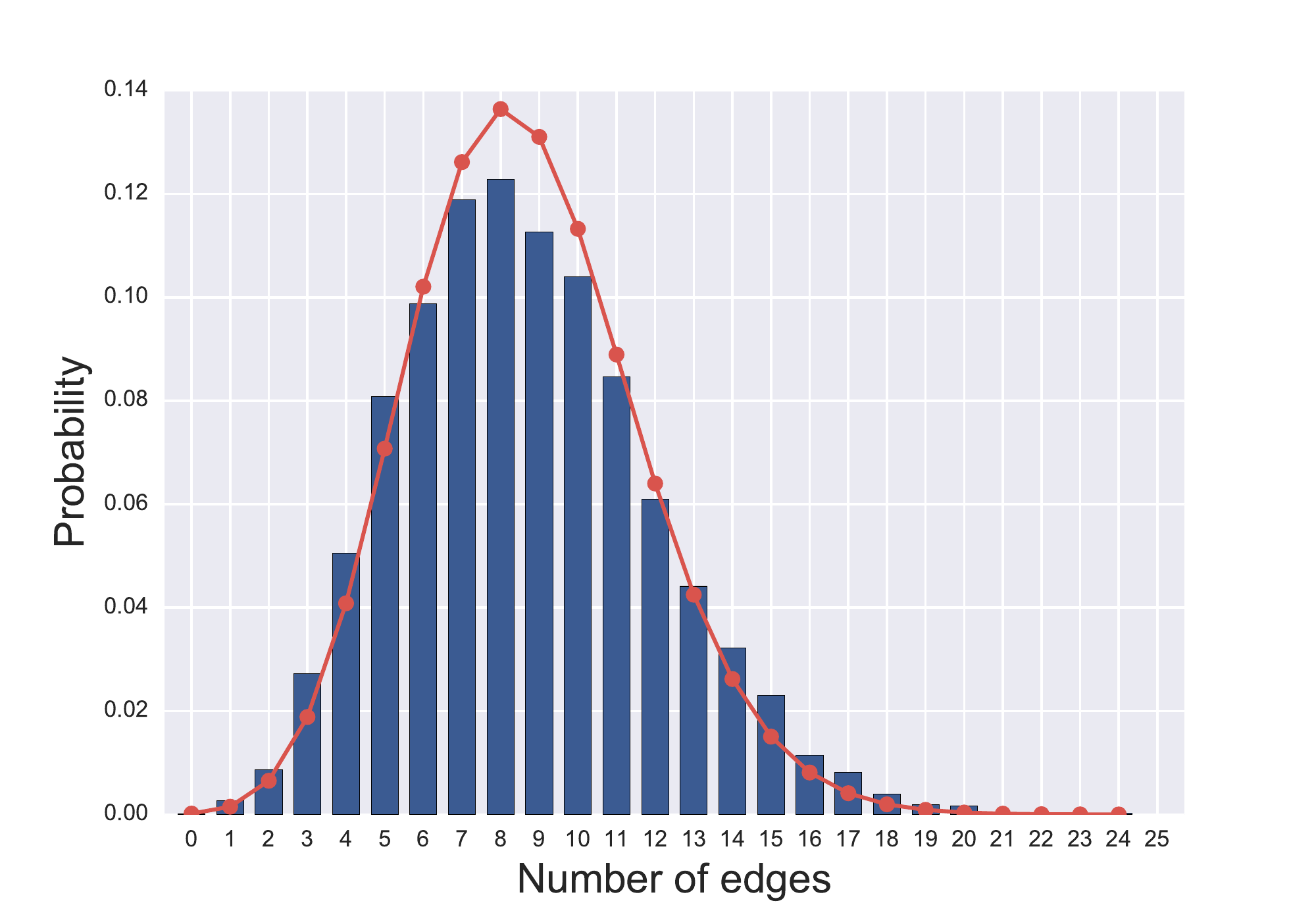}
	\caption{Distribution of the number of edges between two groups of nodes}
	\label{fig:binom_dist}
\end{figure}

\begin{definition}
	\textit{An attribute association $\Delta_{\vec{a_1}, \vec{a_2}}$ between $C_1$ and $C_2$ is statistically significant if the probability that the observed or more number of edges between $C_1$ and $C_2$ is less than $\alpha$ which is called a significance level.}
\end{definition}

In order to show the assumption that the number of edges between two groups of nodes follows the binomial distribution is reasonable, we randomly sampled two groups of 50 nodes from the \textit{DBLP co-authorship network} (the details of the network is described in Section~\ref{sec:experiments}) 10,000 times and obtained the empirical distribution of the number of edges residing between the two groups. As shown in \figurename~\ref{fig:binom_dist}, the empirical distribution (blue bar, mean: 8.68 / stddev: 3.29) is very closed to the actual binomial distribution (red line, mean: 8.64 / stddev: 2.93), which is verified by the chi-squared testing on the two distributions.

\subsection{Locality Preserving Significant Associations}
\label{sec:locality_preserving_signigicant_associations}
An attribute association may reside in anywhere over the entire graph $G$. However, we expect that a certain attribute association could be observed more frequently among nodes which are closed to each other. For example, in the \textit{DBLP co-authorship network}, some authors who have published papers in venues of data mining area are expected to have a certain attribute association with other authors in the same or similar area (e.g., the association of \{\texttt{ICDM}, \texttt{KDD}\} - \{\texttt{ICDM}, \texttt{NIPS}, \texttt{ICML}\}). Any pair of authors in a relationship with the association could be seen in several locations of $G$, but some of them may be located very closely in terms of the hop distance in the graph and form a densely connected subgraph or community. Different communties that have the same venue pattern many times would be corresponding to different schools in different countries. That is, some attribute association patterns come with locality in the graph and such a pattern can be more statistically significant locally rather than globally. Besides, some attribute association patterns that are statistically significant locally may form another complex patterns (e.g., star or chain, not just pair) among them. One of the nice features of the algorithm we propose in Section~\ref{sec:graph_transformation} is that it is able to effectively find all the statistically significant attribute associations while preserving the locality.

\section{Graph Transformation}
\label{sec:graph_transformation}
In this section, we describe the algorithm that finds statistically significant attribute associations in a given attribute graph $G$. The basic approach for finding statistically significant attribute associations is to transform the original graph $G$ into a new graph $\mathcal{AG}=(\mathcal{V},\mathcal{E},\mathcal{A})$, which is called \textit{Association Graph}, where each node in $\mathcal{V}$ corresponds to a group of nodes in $V$ which have the same or similar attribute values, each edge in $\mathcal{E}$ is an attribute association $\Delta$ between two attribute vectors, each attribute vector in $\mathcal{A}$ represents one shared by a group of nodes in $\mathcal{V}$. To avoid confusion, from now on we call a node in $\mathcal{V}$ a cluster and call an edge in $\mathcal{E}$ an association. Each association $\Delta$ is assigned a weight, referred as its strength $w(\Delta)$, that is given by the number of edges between nodes in the clusters forming the association. For a given association $\Delta$ and its associated strength, defined as the number of edges between nodes in the clusters, we can determine whether $\Delta$ is significant or not by looking at the strength and the size of the clusters to which $\Delta$ is incident, which will be explained in detail in the following sections.

\begin{figure}[t]
	\centering
	\includegraphics[width=0.4\textwidth]{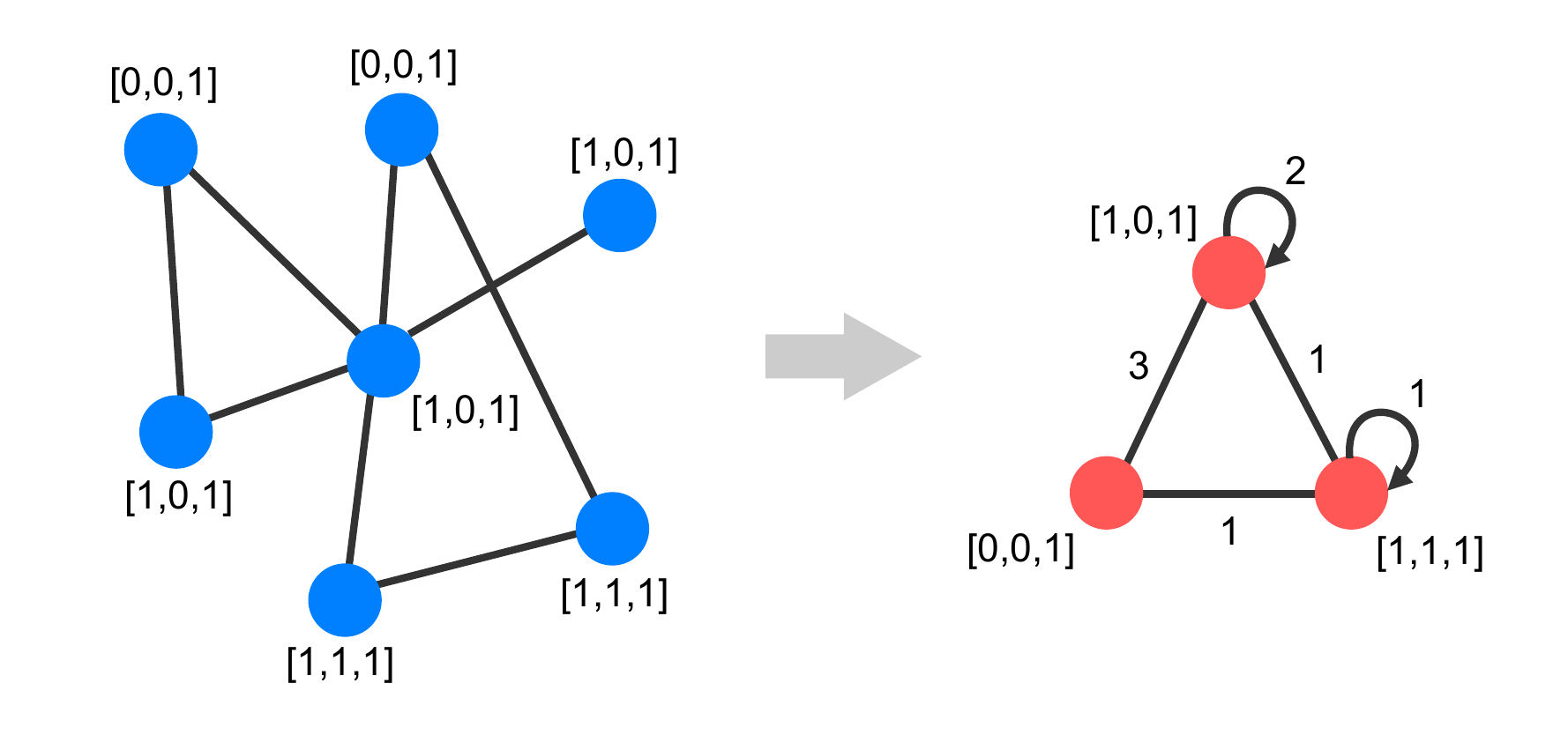}
	\caption{Graph transformation}
	\label{fig:graph_transform}
\end{figure}

The graph transformation can be done through an iteration of two steps. We first start with a single cluster that contains all nodes of $V$ in $G$ and then the cluster is partitioned into several subclusters by applying two steps repeatedly and iteratively. For the first step, a cluster is split such that each subcluster contains a subset of $V$ that have similar attribute values. This operation is able to be easily done using any clustering algorithms. In case of binary attributes, we just select one of the attributes and then do two-way split with respect to the attribute. In Section~\ref{sec:similarity-based_split}, we explain how to select the attribute. For the second step, we try to split a cluster such that each of the associations incident to the cluster has higher strength in order to obtain more significant associations between two sets of attributes. That is, the iteration of the two different splits alternate between performing the similarity-based split, which produces clusters with the same or similar attribute values, and the strength-based split, which makes associations more significant. It results in a new graph $\mathcal{AG}$ where we can see groups of nodes with certain attribute values and significant associations between them, as shown in \figurename~\ref{fig:graph_transform}.

Algorithm~\ref{alg:graph_transformation} shows the whole structure of the graph tranformation algorithm including the two steps of splits. and the following subsections describe how each split should be done in detail.

\subsection{Similarity-based split}
\label{sec:similarity-based_split}
As mentioned already, the goal of the first step is to maximize the similarity among attribute values in each cluster so that each cluster can represent a certain set of attribute values. Thus we select one of the clusters in $\mathcal{AG}$ and then split it into two subclusters based on a certain attribute so that each subcluster contains a set of node that share the same value on the attribute. The way to select a cluster in $\mathcal{AG}$ is based on the following idea. Basically we do not only want to maximize the similarity of attribute values in each subcluster after the split, but also want each subcluster to be statistically significant as much as possible in terms of the attribute values of its nodes.

\begin{algorithm}[t]
	\caption{Algorithm for graph transformation}
	\label{alg:graph_transformation}
	\begin{algorithmic}[1]
		\renewcommand{\algorithmicrequire}{\textbf{Input:}}
		\renewcommand{\algorithmicensure}{\textbf{Output:}}
		\REQUIRE $G=(V, E, A)$
		\ENSURE $\mathcal{AG = (V, E, A)}$
		\\ \textit{Initialization} :
		\STATE $\mathcal{V} = \emptyset, \mathcal{E} = \emptyset$
		\STATE $c$ is initialized as a cluster containing all nodes in $V$
		\STATE $\mathcal{V} = \mathcal{V} \cup c$
		\\ \textit{Iterative Process}
		\WHILE {there exist at least one cluster to split}
		\STATE $c = $ \textit{findClusterForSimilaritySplit($\mathcal{AG}$)}
		\IF {($c$ exists)}
		\STATE similaritySplit($\mathcal{AG}$, $c$)
		\ENDIF
		\STATE $c = $ \textit{findClusterForStrengthSplit($\mathcal{AG}$)}
		\IF {($c$ exists)}
		\STATE strengthSplit($\mathcal{AG}$, $c$)
		\ENDIF
		\ENDWHILE
		\RETURN $\mathcal{AG}$
	\end{algorithmic}
\end{algorithm}

To achieve the goal, we need to figure out which cluster should be split and which attribute should be used to split the cluster. Let $p_i$ denote the probability that a value of 1 occurs at $i$-th attribute, which is the fraction of nodes with a value of 1 for the $i$-th attribute in $G$. So, $p_i$ is considered an expectation of having the attribute value for a random node. First, an attribute on which a cluster should be split based is picked such that the probability of the attribute having the value of 1 in the cluster is least deviated from its corresponding $p_i$. It allows the subclusters to not only have higher similar attribute values among the nodes in them but also have the highest significance gain through the split. Once we decide which attribute should be used for the split of the clusters, we select one of the clusters to split. While assuming that the attributes are independent of each other and the number of times the value of 1 appears at the $i$-th attribute follows the binomial distribution with the probability $p_i$, the statistical significance $\Psi_{c}$ of a cluster $c$ is defined based on the product of p-values of the attribute values of the nodes in the cluster as follows,
\begin{equation}
	\Psi_{c} = 1 - \prod_{i=1}^{l} \Big(1 - \sum_{j=0}^{k_i - 1} \binom{|c|}{j} p_i^j (1 - p_i)^{|c| - j} \Big)
\end{equation}

\noindent where $k_i$ is the number of nodes having the value of 1 on the $i$-th attribute and $|c|$ is the number of nodes in the cluster $c$. So for each cluster $c$ we compute $\Psi_{c'}$ of the subclusters $c'$. Remind that our goal is to split a cluster so that its subclusters are most statistically significant. However, since the subclusters may have different significances (one can be highly significant but the others can be very low), we take subclusters with the lowest significance from each of the clusters in $\mathcal{AG}$ and then select a cluster that will produce a subcluster with the highest significance among those subclusters, i.e.,

\begin{eqnarray}
	\argmax_{c} \big(\min_{c' \in sb(c)} \Psi_{c'}\big)
\end{eqnarray}

\noindent where $sb(c)$ is a set of subclusters that will be created after the split. In this way, we can avoid to split a cluster that will produce the least significant subclusters. By repeating this kind of split, $\mathcal{AG}$ will have only clusters, in each of which the same attribute values are shared by its nodes, but we need to place one constraint while doing the split. Even though a cluster represents a certain set of attribute values shared in it, if it contains only a few nodes then its attribute values may not be meaningful at all when we look at an attribute association between clusters in $\mathcal{AG}$. Thus, we use \textit{size\_support}, denoted by $\lambda$, to force a cluster not to split any more if all the subclusters that will be obtained after splitting the cluster have the sizes less than $\lambda \cdot |V|$. Thus, during the first step, we examine only clusters satisfying the $\lambda$ threshold to determine which cluster should be split. Also, it is obvious that a cluster in which all its nodes have the same attribute values does not need to be split. 

We do not only want nodes in the same cluster to have the same attribute values but also allow them to have similar attribute values. In other words, even though every node in a cluster does not agree on a certain attribute, if the distribution of the values of the attribute is statistically significantly deviated from the expectation, then those nodes are considered to have an identical value for the attribute.

Once a cluster is split at the first step, we move on to the second step to increase the significances of the attribute associations between clusters.

\subsection{Strength-based split}
\begin{figure}
	\centering
	\begin{subfigure}[b]{0.4\textwidth}
		\includegraphics[width=\textwidth]{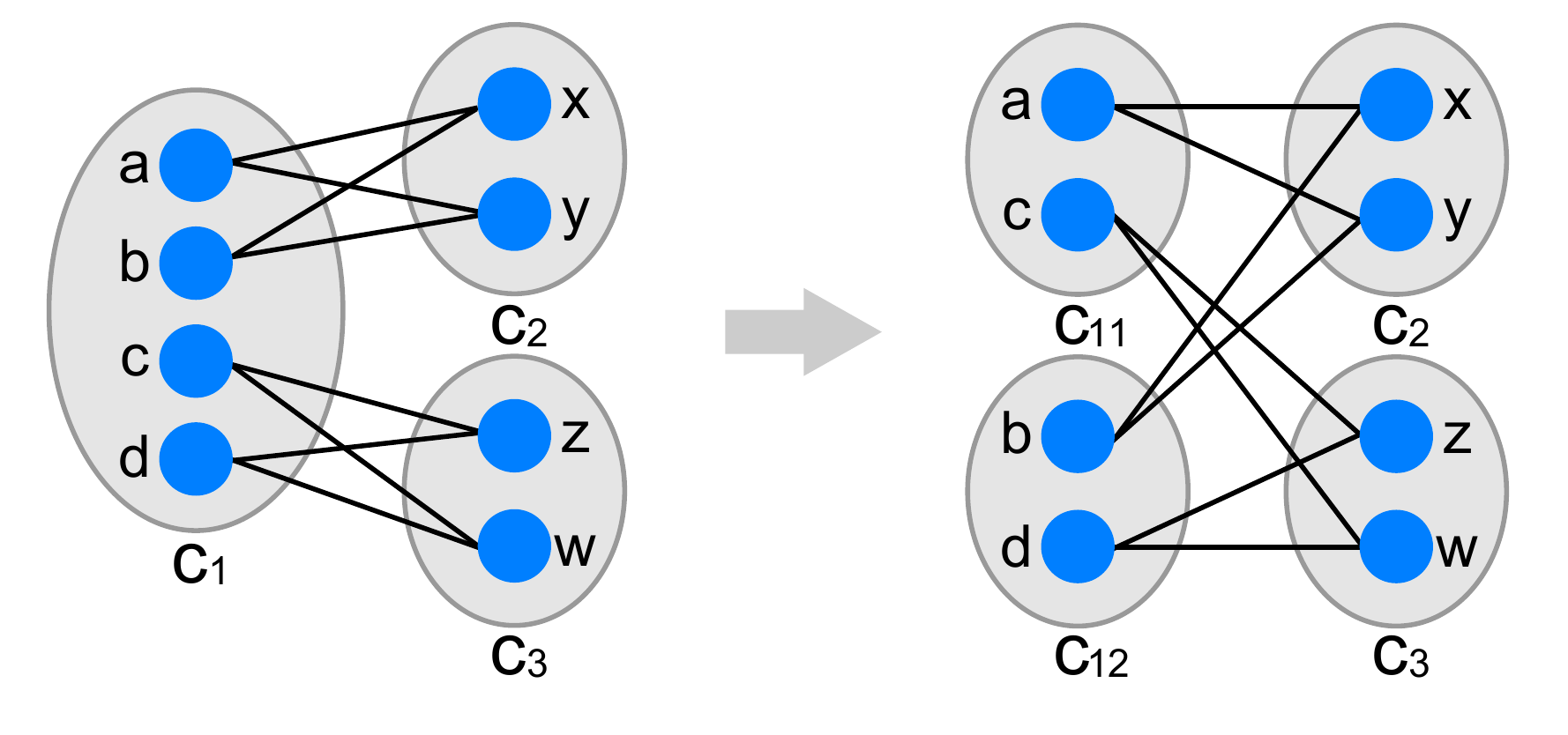}
		\caption{($a$, $c$) and ($b$, $d$)}
        \label{fig:str_split_example_1}
	\end{subfigure}
	
	\begin{subfigure}[b]{0.4\textwidth}
		\includegraphics[width=\textwidth]{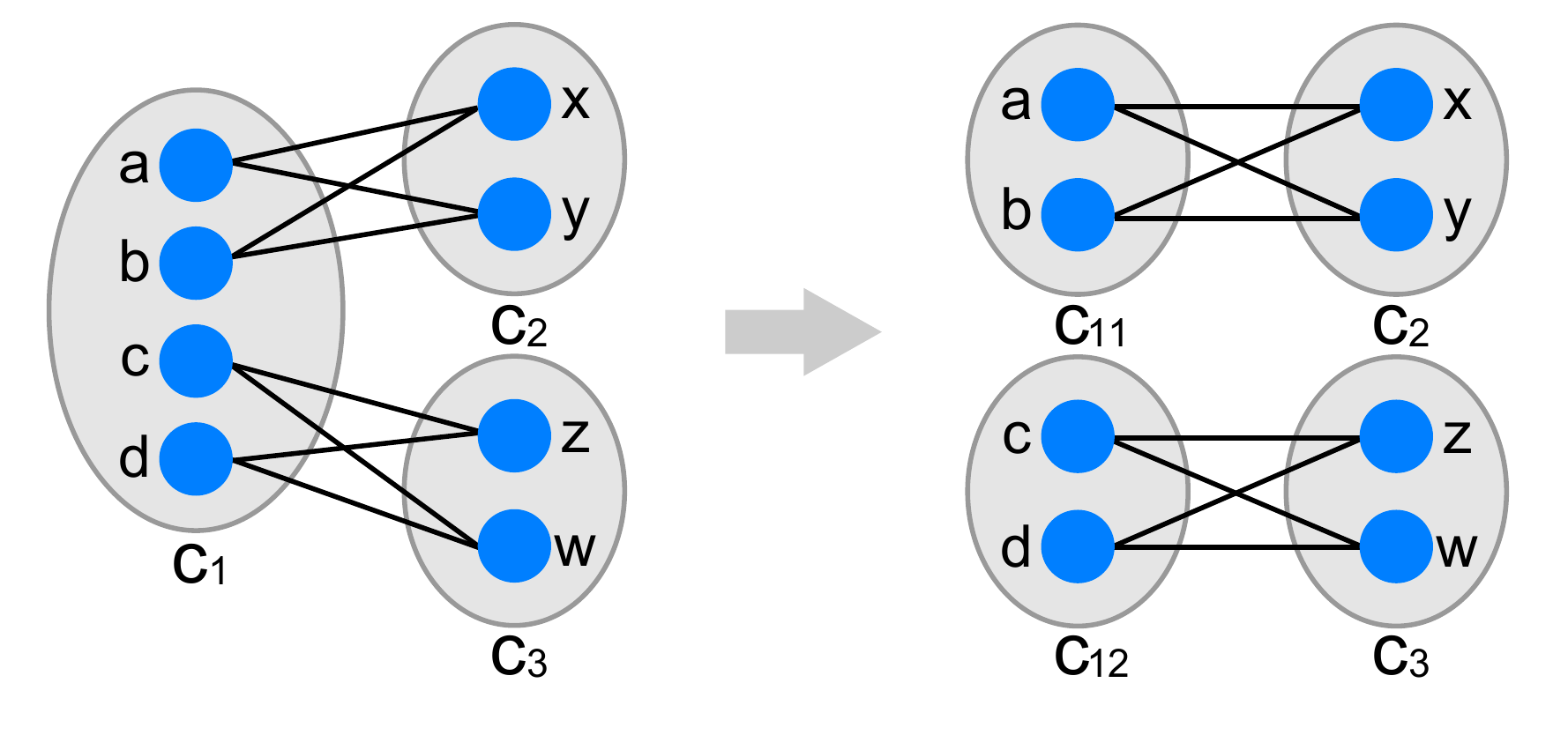}
		\caption{($a$, $b$) and ($c$, $d$)}
        \label{fig:str_split_example_2}
	\end{subfigure}
	\caption{Two different strength-based splits}
	\label{fig:str_split_example}
\end{figure}

While the similarity-based split of the first step aims to increase the similarity of attribute values for a cluster, we try to maximize strengths of associations to which a cluster is incident through the strength-based split. Given an attribute association between two clusters, its strength is defined as the number of edges that connect the nodes of the clusters. The strength is not meaningful by itself because the significance depends on the sizes of the clusters as well as the strength. As we discussed the definition of a statistically significant attribute association in Section~\ref{sec:statistically_significance}, the stronger strength an attribute association has and the smaller the associated clusters are, the higher statistically significant the association is. Thus, in order to make an association more significant, a cluster that is one of the end points of the association needs to be split into subclusters such that nodes which have many common neighbor clusters belong to the same subcluster. \figurename~\ref{fig:str_split_example} illustrates the basic idea of the strength-based split. Suppose we want to maximize the significance of the associations held by the cluster $c_1$ and we consider two different splits to do that as presented in \figurename~\ref{fig:str_split_example_1} and \figurename~\ref{fig:str_split_example_2}. The nodes $a$ and $b$ in $c_1$ have edges, all of which are incident to other nodes in $c_2$ while the nodes $c$ and $d$ are adjacent to only other nodes in $c_3$. Thus, in order for the subclusters obtained from splitting $c_1$ to have associations of maximized significance, the split should produce two subclusters which contain the two nodes $a$ and $b$, and the other two nodes $c$ and $d$, respectively.

So we need to find the optimal split of a cluster so that its associations become more significant. For a given cluster $c$ we try to split, we build a graph $\tilde{G} = (\tilde{V}, \tilde{E})$ where $\tilde{V} = \{u | u \in c \}$ and $\tilde{E} = \{ (u, v) | u, v \in c \land \exists c' \text{ s.t. } (u, w_1), (u, w_2) \in E \text{ and } w_1, w_2 \in c' \}$, some of which are connected to each other if they have edges with some common neighbor clusters, $\Gamma(c)$. Those edges in $\tilde{E}$ are weighted based on the fraction of edges to common neighbors among all of their edges. Then, we partition the graph $\tilde{G}$ based on the weights of the edges in the graph and the subgraphs resulted from the partition become the subclusters we obtain through the strength-based split. For this task, we need to come up with a proper way to assign weights to the edges. We borrow the idea of tie-strength between individuals in social network. In the social science community, there are many different ways to define the tie-strength of an interpersonal relationship~\cite{gupte2012measuring}, and one widely used measure is the Jaccard index. That is, a tie-strength between two individuals $u$ and $v$ is determined by $|\Gamma(u) \cap \Gamma(v)| / |\Gamma(u) \cup \Gamma(v)|$ where $\Gamma(\cdot)$ is a set of neighbors of a node. In our setting, two nodes $u$ and $v$ in the cluster $c$ may not have common neighbor nodes in $G$ but some of their neighbor nodes may belong to the same neighbor cluster $c'$ in $\mathcal{AG}$. Similarly, when $u$ and $v$ in $c$ are connected to some of the nodes in a common neighbor cluster $c'$ of $c$, there might not be common nodes in $c'$ which are incident to both $u$ and $v$. Thus, we modify the Jaccard index slightly so as to measure the tie-strength between $u$ and $v$ while capturing the common neighbor clusters.

\begin{equation}
	TS(u, v) = \frac{\sum_{c' \in \Gamma(c)} \text{min} \{\phi(u, c'), \phi(v, c')\}}{\sum_{c' \in \Gamma(c)} \text{max} \{\phi(u, c'), \phi(v, c')\}}
\end{equation}

\noindent where $\phi(u, c') = |{w | w \in c \land (u, v) \in E}|$, that is the number of edges in $E$ between $u$ and any nodes in $c'$. Using this tie-strength measure, we can have nodes belong to the same subcluster after the split if they have many common neighbor clusters, regardless of whether they have common neighbor nodes in $G$ or not (of course, it depends on the weight given by $TS(\cdot, \cdot)$).

Once we have $\tilde{G}$ for the cluster $c$ then we perform graph partitioning on $\tilde{G}$ to find optimal subclusters that can make the associations between $c$ and $c' \in \Gamma(c)$ more significant. Since all the edges $\tilde{E}$ of $\tilde{G}$ are assigned weights and $\tilde{G}$ should be partitioned based on the weights, we take an approach to maximize the modularity of $\tilde{G}$~\cite{blondel2008fast}. The modularity $Q(\tilde{G})$ is defined as

\begin{equation}
	Q(\tilde{G}) = \frac{1}{2m} \sum_{u, v} \Big[ A_{uv} - \frac{k_u, k_v}{2m}\Big] \delta(c_u, c_v)
\end{equation}

\noindent where $m = \tilde{E}$, $k_u$ is the degree of $u$, $c_u$ is the group to which $u$ belongs, and $A_{uv}$ is 1 if there is an edge in $\tilde{E}$ between $u$ and $v$ otherwise 0. That is, the modularity is the fraction of the edges that fall within the given groups minus the expected such fraction if edges were distributed at random. If we split the cluster $c$ through the graph partitioning method as described, a set of nodes that share many common neighbor clusters is likely to fall within the same subcluster as much as possible, and different nodes that share only few common neighbors would be distributed to different subclusters. Thus, we can increase the statistical significances of the attribute associations.

During the second step, we enforce a couple of conditions to prune some clusters and associations in $\mathcal{AG}$ and do not perform the strength-based split on them for both achieving computational efficiency and finding more meaningful results. As done in the first step, we use \textit{size\_support}, $\lambda$ because if the size of a cluster $c$ is too small, we do not believe that $c$ is representative of a certain set of attribute values. Thus, the strength-based split is run for a cluster $c$ only when $|c| \geq \lambda \cdot |V|$. In addition to that, if a cluster has an attribute association with too weak strength, then we can safely discard it for the rest of the algorithm. Note that the strength of an attribute association between two clusters monotonically decreases as the two splits are performed iteratively while the statistically significance is not monotonic in either way. Since we consider only attribute associations between clusters satisfying the \textit{size\_support} condition and the statistically significance of an association depends on its strength and the sizes of the clusters at the end points, we can prune an attribute association from $\mathcal{AG}$ as long as it meets the following condition.

\begin{lemma}
	\label{lemma:pruning}
	\textit{Given an attribute association $\Delta_{c_1, c_2}$ and its two incident clusters $c_1$ and $c_2$, if the strength of $\Delta_{c_1, c_2}$ is less than $\Phi^{-1} \Big(1 - \alpha - \frac{C(p^2 + q^2)}{\sqrt{npq}}\Big) \sqrt{npq} + np$, then $\Delta_{c_1, c_2}$ does not have a chance to be statistically significant any more, where $n=|c_1| \cdot |c_2|$, $p=\delta_G$, $q=1-p$, $\Phi(\cdot)$ is the error function, and $C$ is a constant.}
\end{lemma}

\begin{IEEEproof}
	\label{pf:pruning}
	Given the size\_support $\lambda$, both the clusters $c_1$ and $c_2$ should have the size of at least $|V| \cdot \lambda$ in order to make the attribute association $\Delta_{c_1, c_2}$ considered as statistically significant. Also, let $k$ denote the strength of $\Delta_{c_1, c_2}$ and then according to the (\ref{eq:pval_association}), $P[X \geq k] \leq \alpha$. If we approximate the binomial distribution using the normal distribution,
	\setlength{\arraycolsep}{0.0em}
	\begin{eqnarray}
		 P[X \geq k] & {} = {} & P\big[\frac{X-np}{\sqrt{npq}} \geq \frac{k-np}{\sqrt{npq}}\big] \nonumber \\
		 & = & P\big[Z \geq \frac{k-np}{\sqrt{npq}}\big] \leq \alpha
		 \label{eq:noraml_approx}
	\end{eqnarray}
	\setlength{\arraycolsep}{5pt}
	Now we have the standard normal distribution and need to find the lower bound of $k$ which satisfies the inequality~(\ref{eq:noraml_approx}). Using the error function $\Phi(x) = \frac{1}{\sqrt{2\pi}}\int_{-\infty}^{x} e^{\frac{-t^2}{2}}\,\mathrm{d}t$ which is essentially identical to the standard normal cumulative distribution function~\cite{greene2003econometric},
	\begin{eqnarray}
		\frac{k-np}{\sqrt{npq}} \geq \Phi^{-1}(1-\alpha) \nonumber \\
		k \geq \Phi^{-1}(1 - \alpha) \sqrt{npq} + np
		\label{eq:k_lower_bound}
	\end{eqnarray}
	The lower bound for $k$ is originated from approximation based on the standard normal distribution and thus we need to get the error bound. According to the following \textit{Berry-Essen theorem}~\cite{nagaev2012bound},
	\begin{equation}
			\sup_{x \in \mathds{R}} \Bigl| P \Big[\frac{B(p,n) - np}{\sqrt{npq}} - \Phi(x)\Big] \Bigr| \leq \frac{C(p^2 + q^2)}{\sqrt{npq}}
	\end{equation}
		with $C < 0.4748$, we know that the error arising from the approximation is at most $\frac{C(p^2 + q^2)}{\sqrt{npq}}$. As a result, if we relax the lower bound for $k$ in (\ref{eq:k_lower_bound}) to the extent of the error, then we obtain
	\begin{equation}
		k \geq \Phi^{-1}\Big(1 - \alpha - \frac{C(p^2 + q^2)}{\sqrt{npq}}\Big) \sqrt{npq} + np
	\end{equation}
\end{IEEEproof}

Since $p$ is very small and $n$ is large for given $\lambda$, the error bound is small and we can still get a reasonably tight lower bound for $k$. Regarding the inverse of the error function, if we use $\alpha=0.01$ as the significance level, then $\Phi^{-1} (1 - \alpha) = 1.8212$. According to Lemma~\ref{lemma:pruning}, we drop attribute associations if they are too weak to be able to be significant later on. In fact, such associations are noise and do not bring us any meanings. Rather, it prevents the strength-based split from running optimally.

\section{Experiments}
\label{sec:experiments}

\begin{table}[t]
	\centering
	\begin{tabular}{|c|c|c|c||c|c|}
		\cline{2-6}
		\multicolumn{1}{r|}{} & \multicolumn{3}{|c||}{Original graph} & \multicolumn{2}{|c|}{Association graph} \\
		\cline{2-6}
		\multicolumn{1}{r|}{} & Nodes & Edges & Density & Nodes & Edges \\
		\cline{2-6}
		\hline
		DBLP & 4,672 & 37,726 & 0.00346 & 195 & 6,302 \\ \hline
		Yelp & 4,454 & 44,906 & 0.00453 & 202 & 8,388 \\
		\hline
	\end{tabular}
	\caption{Dataset statistics}
	\label{tab:dataset_stats}
\end{table}
\subsection{Datasets}
\begin{table}[t]
	\centering
	\begin{tabular}{|p{10mm}|p{70mm}|}
		\hline
		Subarea & Conferences \\
		\hline
		DM/ML & ICDM, NIPS, ICML \\ \hline
		OS & SOSP, OSDI \\ \hline
		Theory & FOCS, STOC, SODA \\ \hline
		Security & IEEE Symposium on Security and Privacy (S\&P), ACM Conference on Computer and Communications Security (CCS) \\
		\hline
	\end{tabular}
	\caption{DBLP subareas in computer science}
	\label{tab:DBLP_subareas}
\end{table}

We ran the graph transformation algorithm on real co-authorship and social networks, and obtained the resulting association graphs. Using the association graphs, we analyzed qualitative differences between the statistically significant and frequent associations.
Also we showed the application of the significant patterns to a link prediction problem, and synthetic graphs with attributes are considered to show the algorithm's scalability.

\textbf{DBLP.}
We obtained a collection of bibliographic information from the DBLP website \cite{DBLPwebsite}, an open bibliographic information provider of computer science journals and conferences. Each record of journal or conference paper has one or more authors, and the venue, on which it is published. We first filtered out any authors who appear in less than 3 papers. Then, we considered only papers published to the 10 conferences of 4 different subareas of computer science, i.e., data mining and machine learning (DM/ML), operating systems (OS), theory, and security. More details are shown in Table~\ref{tab:DBLP_subareas}. Then we built an attribute vector of length 10 for each node, i.e., if an author (or a node) published a paper to a conference in Table~\ref{tab:DBLP_subareas}, we set the corresponding vector value to 1. If not, we set the corresponding vector value to 0. Finally, an edge is formed if two authors (or nodes) have co-authored at least one paper in the dataset.

\textbf{Yelp.}
Yelp is a provider of crowd-sourced reviews about local businesses, along with a social network. The Yelp challenge dataset \cite{YelpChallenge} contains the social network, composed of the users (nodes) and their friend relations (edges). Also the sets of users' reviews are provided in the dataset. Each review is tied to a user and a business, and each business has a small set of business type categories. We first filtered out any users who have less than 10 reviews. Then, we considered only reviews for the restaurants, which has at least one of the 10 business categories, \{Chinese, Japanese, Mediterranean, Thai, French, Greek, Vietnamese, Korean, Indian, British\}. The node attributes are compiled similarly to the DBLP dataset. Note that we did not use some of the most popular restaurant categories, e.g., American, Mexican, and Italian. As the majority of users has left reviews on the restaurants of such categories, they seem to appear in most of the attribute associations and carry little or no information.

\begin{figure}[!t]
	\centering
	\begin{subfigure}{\columnwidth}
		\centering
		\begin{tikzpicture}[scale=0.8]
			\begin{axis}[
				ylabel={Density},
				ymin=0, ymax=0.03,
				xtick=data,
				ytick={0, 0.005, 0.01, 0.015, 0.02, 0.025, 0.03},
				yticklabel style={
				            /pgf/number format/fixed,
				            /pgf/number format/precision=3,
				            /pgf/number format/fixed zerofill
				        },
				        scaled y ticks=false,
				ymajorgrids=true,
				axis y line=left,
				grid style=dashed,
				symbolic x coords={DM/ML,OS,Theory,Security},
				cycle list name=exotic,
				every axis plot/.append style={area legend}				
			]
			\addplot [ybar, draw=blue, pattern=horizontal lines dark blue, xshift=-0.5*\pgfplotbarwidth,
				legend image post style={xshift=0.5*\pgfplotbarwidth}
			] table[x=topic, y=density, col sep=comma] {chart/DBLP_subgraph_density.csv}; \label{subfig:subgraph_density}
			\end{axis}
			\begin{axis}[
				xtick style={draw=none}
				ylabel={Number of nodes},
				ymin=0, ymax=3000,
				ytick={0, 500, 1000, 1500, 2000, 2500, 3000},
				axis y line=right,
				axis x line=none,
				symbolic x coords={DM/ML,OS,Theory,Security},
				cycle list name=exotic,
				every axis plot/.append style={area legend}				
			]
			\addlegendimage{draw=blue, pattern=horizontal lines dark blue}
			\addlegendentry{Density}
			\addplot [ybar, draw=blue, pattern=horizontal lines light blue, xshift=0.5*\pgfplotbarwidth,
				legend image post style={xshift=-0.5*\pgfplotbarwidth}
			] table[x=topic, y=nodes, col sep=comma] {chart/DBLP_subgraph_density.csv};
			\addlegendentry{\# of nodes}
			\end{axis}
		\end{tikzpicture}
	\end{subfigure}
	~
	\begin{subfigure}{\columnwidth}
		\centering
		\begin{tabular}{|c|c|c|c|}
			\hline
			Subarea & Nodes & Edges & Density \\
			\hline
			\hline
			DM/ML &  2,657 & 15,639 & 0.00443 \\ \hline
			OS & 395 & 2,113 & 0.02715 \\ \hline
			Theory & 1,700 & 16,738 & 0.01159 \\ \hline
			Security & 758 & 4,291 & 0.01496 \\
			\hline
		\end{tabular}
	\end{subfigure}
	\caption{DBLP graph statistics for different subareas}
	\label{fig:DBLP_subgraph_stats}
\end{figure}
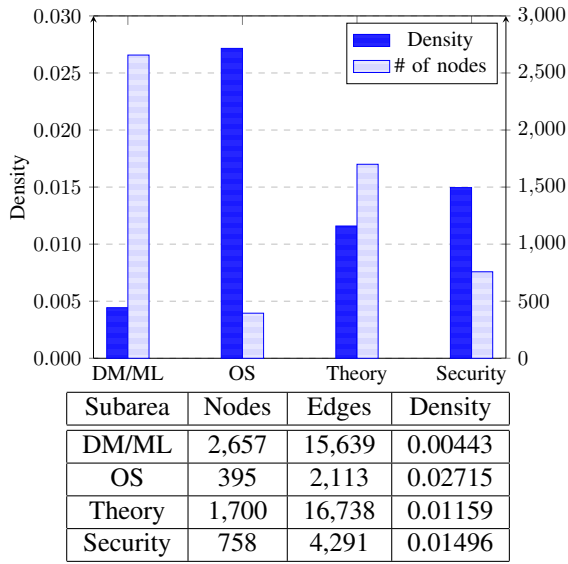

\subsection{Effectiveness Analysis}
To evaluate the effectiveness of our algorithm, we conducted the set difference between the statistically significant associations and the top-$15$ frequent associations. As the resulting significant associations contain wildcard attributes, it is not easy to make direct comparisons or set differences between the two. Thus, we took a conservative approach that as long as all attribute values of any top-$15$ frequent associations have exact or wildcard attribute match, we considered that there is a match. This approach is certainly in favor of the frequent associations, since it ignores that wildcard matches may lead to some other possible set of attribute values.

\begin{table}[t]
	\centering
	\begin{tabular}{|c|rcl|}
\hline
\# & \multicolumn{3}{|c|}{Association} \\
\hline
1 & \{SOSP, OSDI, S\&P, CCS\}  &--&  \{SOSP, OSDI\} \\ \hline
2 & \{SOSP, OSDI, S\&P, CCS\}  &--&  \{S\&P, CCS\} \\ \hline
3 & \{ICML, ICDM, S\&P(*)\}    &--&  \{ICML, ICDM\} \\ \hline
4 & \{SOSP, OSDI, S\&P, CCS\}  &--&  \{SOSP, S\&P, CCS\} \\ \hline
5 & \{FOCS(*), STOC(*), CCS\} &--&  \{S\&P, CCS\} \\ \hline
6 & \{ICML, ICDM, S\&P(*)\}    &--&  \{ICML(*), ICDM\} \\ \hline
7 & \{SOSP, S\&P, CCS\}        &--&  \{SOSP, OSDI\} \\ \hline
8 & \{SOSP, S\&P, CCS\}        &--&  \{S\&P, CCS\} \\ \hline
9 & \{ICML, ICDM, S\&P(*)\}    &--&  \{ICDM, OSDI(*)\} \\ \hline
10 & \{ICML, ICDM\}            &--&  \{ICML(*), ICDM\} \\ 
\hline
	\end{tabular}
	\caption{Significant associations minus Frequent associations for DBLP}
	\label{tab:DBLP_sig_association_result}
\end{table}

Table~\ref{tab:DBLP_sig_association_result}, we have the set difference between statistically significant and frequent associations for the DBLP dataset. First consider 4 subgraphs that only contains the nodes and their edges, whose attribute value for any conference of the corresponding subarea is 1. \figurename~\ref{fig:DBLP_subgraph_stats} describes the characteristics of each subgraph. The subgraph of DM/ML has a large number of nodes but its graph density is small, which means that the tie-strengths are weak. On the other hand, there are relatively small numbers of nodes in the subgraph of OS and security but their densities are high, which means that the tie-strengths are strong among the nodes. We can easily identify the OS and security-related associations, which contain \{SOSP, OSDI\} and \{S\&P, CCS\}, are appearing on top of the difference list. Also note that many frequent associations are related to DM/ML conferences since its subgraph contains the most number of edges while its density is low.

From Table~\ref{tab:DBLP_sig_association_result}, we can infer many interesting significant associations, which do not appear in the frequent association list. The association number 1, 2, 4, 7, and 8 clearly shows that the nodes who have authorship in the OS-related conferences tend to co-work with the authors in the security-related conferences. The association number 3, 5 and 6 shows that the nodes who have authorship in the security-related conferences frequently co-work with the authors in DM/ML and theory-related conferences. Interestingly enough, the association number 9 shows how the authors in DM/ML, security, and OS have frequent co-authorship relations in the graph. These results might look obvious to some of the readers who have a good understanding of co-authorship in computer science. However, when the relationships of attributes are little known, the discussed results may be intriguing.

\begin{table*}[t]
	\centering
	\begin{tabular}{|@{\hspace{1mm}}c@{\hspace{1mm}}|r@{\hspace{1mm}}c@{\hspace{1mm}}l|}
\hline
\# & \multicolumn{3}{|c|}{Association} \\
\hline
  1 & \{Chinese, Japanese, Mediterranean, Thai, Greek\} &--&  \{Chinese, Mediterranean, Thai(*), Greek\} \\ \hline
  2 & \{Chinese, Japanese, Mediterranean, Thai, Vietnamese, Korean\}  &--&  \{Chinese, Japanese, Mediterranean, Thai, Greek\} \\ \hline
  3 & \{Chinese, Japanese, Thai, Vietnamese, Korean\}  &--&  \{Chinese, Japanese, Vietnamese, Korean\} \\ \hline
  4 & \{Chinese, Japanese, Mediterranean, Thai, Vietnamese, Korean, Indian\}  &--&  \{Chinese, Japanese, Mediterranean, Thai, Vietnamese, Korean\} \\ \hline
  5 & \{Chinese, Mediterranean, Thai(*), Greek\}  &--&  \{Chinese, Mediterranean\} \\ \hline
  6 & \{Chinese, Japanese, Mediterranean, Thai, Vietnamese, Korean, Indian\}  &--&  \{Chinese, Japanese, Thai\} \\ \hline
  7 & \{Chinese, Japanese, Mediterranean, Thai, Vietnamese, Korean\}  &--&  \{Chinese, Mediterranean, Thai(*), Greek\} \\ \hline
  8 & \{Chinese, Japanese, Mediterranean, Thai, Greek\}  &--&  \{Chinese, Mediterranean\} \\ \hline
  9 & \{Chinese, Japanese, Thai, Vietnamese, Korean\}  &--&  \{Chinese, Japanese, Thai\} \\ \hline
 10 & \{Chinese, Japanese, Mediterranean, Thai, Vietnamese, Korean, Indian\}  &--&  \{Chinese, Japanese, Mediterranean, Thai, Greek\} \\
\hline
	\end{tabular}
	\caption{Significant associations minus Frequent associations for Yelp}
	\label{tab:Yelp_sig_association_result}
\end{table*}

Table~\ref{tab:Yelp_sig_association_result} shows that the set difference between statistically significant and frequent associations for the Yelp dataset. Note that \{Chinese, Japanese\} appears very commonly in the association results due to their prevalence in node attributes. Thus, we will exclude them from the subsequent discussions. Also it turned out that the first 10 significant associations with the highest statistical significance are the same as the associations reported in Table~\ref{tab:Yelp_sig_association_result}. That is, none of the first 10 significant associations are reported in the top-$15$ frequent association results, since the significant associations do not occur often in terms of frequency but do occur often in the dataset in a statistically significant manner.

Among the frequent visitors of \{Mediterranean, Thai\}, the association number 2 and 7 shows that the nodes with \{Greek\} attribute are strongly associated with the nodes with \{Vietnamese, Korean\} attributes, and the association number 4, 6 and 10 shows that the nodes with \{Vietnamese, Korean, Indian\} are strongly associated with the nodes with \{Vietnamese, Korean\} and \{Greek\} attributes. Also the association number 5 and 8 describes that the nodes with \{Mediterranean\} have statistically significant associations with the nodes with \{Mediterranean, Thai, Greek\}.

\subsection{Scalability Analysis}
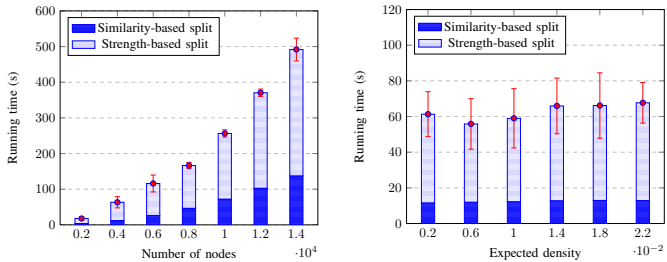
\begin{figure}[!t]
	\centering
	\begin{subfigure}{0.48\columnwidth}
		\centering
		\begin{tikzpicture}[scale=0.5]
			\begin{axis}[
				ybar stacked,
				xlabel={Number of nodes},
				ylabel={Running time (s)},
				ymin=0, ymax=600,
				xtick=data,
				ytick={0, 100, 200, 300, 400, 500, 600},
				ymajorgrids=true,
				grid style=dashed,
				legend pos=north west,
				cycle list name=exotic,
				every axis plot/.append style={area legend}
			]
			\addplot [draw=blue, pattern=horizontal lines dark blue]
				table[x=nnode, y=avg_node_split, col sep=comma] {chart/summary_scalability_nodes.csv};
			\addlegendentry{Similarity-based split}
			\addplot [draw=blue, pattern=horizontal lines light blue, mark=*, mark size=2.0, mark options={fill=red}, error bars/y dir=both, error bars/y explicit, error bars/error bar style={color=red}]
				table[x=nnode, y=avg_edge_split, y error=std_total, col sep=comma] {chart/summary_scalability_nodes.csv};
			\addlegendentry{Strength-based split}
			\end{axis}
		\end{tikzpicture}
		\caption{$\mu=0.6$, density$=0.010$}
		\label{fig:scalability_nnodes_vs_runtime}
	\end{subfigure}
	~
	\begin{subfigure}{0.48\columnwidth}
		\centering
		\begin{tikzpicture}[scale=0.5]
			\begin{axis}[
				ybar stacked,
				xlabel={Expected density},
				ylabel={Running time (s)},
				ymin=0, ymax=120,
				xtick=data,
				ytick={0, 20, 40, 60, 80, 100, 120},
				ymajorgrids=true,
				grid style=dashed,
				legend pos=north west,
				cycle list name=exotic,
				every axis plot/.append style={area legend}
			]
			\addplot [draw=blue, pattern=horizontal lines dark blue]
				table[x=expected_density, y=avg_node_split, col sep=comma] {chart/summary_scalability_densities.csv};
			\addlegendentry{Similarity-based split}
			\addplot [draw=blue, pattern=horizontal lines light blue, mark=*, mark size=2.0, mark options={fill=red}, error bars/y dir=both, error bars/y explicit, error bars/error bar style={color=red}]
				table[x=expected_density, y=avg_edge_split, y error=std_total, col sep=comma] {chart/summary_scalability_densities.csv};
			\addlegendentry{Strength-based split}
			\end{axis}
		\end{tikzpicture}
		\caption{$\mu=0.6$, nodes$=4,000$}
		\label{fig:scalability_densities_vs_runtime}
	\end{subfigure}
	\caption{Running time experiments on synthetic graph datasets}
	\label{fig:scalability_experiments}
\end{figure}

We evaluated the computation cost of our algorithm on synthetic attributed graphs of different sizes and densities. The experiments were carried on a machine with an Intel Xeon 3.1GHz CPU and 32GB memory, running 64bit Ubuntu 14.04. All algorithms are implemented in Python 2.7.

The graphs are generated based on the simplified version of Multiplicative Attribute Graph (MAG) model \cite{kim2012multiplicative}. MAG is widely used in the literature to generate synthetic graphs with node attributes, and known to model real-world networks with flexibility. We conducted two sets of experiments with $l=5$ and $\mu$'s fixed, the probability of each attribute value being 1, i.e., each node has five binary attributes and the attributes are drawn from the same distribution, retaining the node attribute distribution throughout the experiments.

\textbf{Time complexity.}
Our algorithm is divisive in nature and it splits at least one node of \textit{Association Graph} in every iteration. First, the similarity-based split step will run $\mathcal{O}(2^l)$ iterations. Usually the length of attribute vector is small, $l \ll n$, and the similarity-based split under reasonable settings takes much less time compared to that of the strength-based split. In the strength-based split step, it is not hard to see that the computation of tie-strengths between each pair of nodes, $\mathcal{O}(n^2)$, dominates the running time of the step. And we can notice that the algorithm will run $\log n$ iterations of the strength-based steps on average. Accordingly the overall average time complexity of the algorithm is $\mathcal{O}(n^2 \log n)$.

\textbf{Results.}
\figurename~\ref{fig:scalability_nnodes_vs_runtime} shows the computation time over the number of nodes. We fixed the attribute link-affinity matrix \cite{kim2012multiplicative}, which determines the probability of edge formation between two sets of node attributes. Note that since we kept all parameters of the MAG model but the number of nodes, the graph density remained the same. We confirmed that the algorithm is of polynomial time in the number of nodes. This result is in line with the time complexity we discussed above.

In the second experiment, we fixed the number of nodes and the scale factor of the attribute matrix, which merely changes the expected number of edges. That is, we scaled the attribute matrix such that the resulting graphs have the graph densities as we desire, without changing any other properties of the graphs. In \figurename~\ref{fig:scalability_densities_vs_runtime}, we can easily observe that the algorithm's running time remains almost the same as we increase the expected graph density. The aforementioned time complexity should well explain the result.

Finally, both of the plots in \figurename~\ref{fig:scalability_experiments} show that the running time of the strength-based split step dominates that of the similarity-based step. Also both plots describe that the running time of the similarity-based step remain the same as we add more edges with the number of nodes fixed, and the running time grows as we increase the number of nodes. This supports our intuition that the similarity-based split step is not relevant to the number of edges or graph density.

\subsection{Application: Link prediction}
\label{sec:link_prediction}

\begin{figure}[t]
	\centering
	\begin{tikzpicture}[scale=0.7]
		\begin{axis}[
				xlabel={False positive rate},
				ylabel={True positive rate},
				xmin=0, xmax=1,
				ymin=0, ymax=1,
				xtick={0.2,0.4,0.6,0.8,1.0},
			    ytick={0.2,0.4,0.6,0.8,1.0},
				ymajorgrids=true,
				grid style=dashed,
				width=0.4\textwidth,
				legend style={legend pos=south east,font=\small},
				legend entries={Jaccard+Significant, Jaccard+Freqeunt},
			]
			\addplot[color=blue, mark=x] table[x=jacc_sig_fpr, y=jacc_sig_tpr, col sep=comma] {chart/prediction_result_2_roc.csv};
			\addplot[color=red, mark=o] table[x=jacc_freq_fpr, y=jacc_freq_tpr, col sep=comma] {chart/prediction_result_2_roc.csv};
		\end{axis}
	\end{tikzpicture}
	\caption{Link prediction performance}
	\label{fig:link_prediction}
\end{figure}
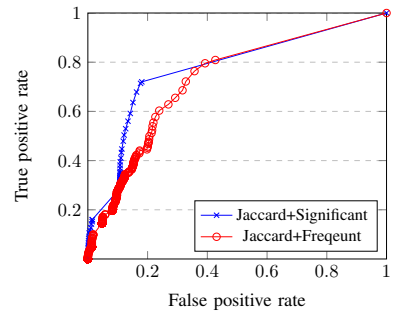

As one of the application for which the statistically significant attribute associations are useful, the \textit{link prediction} problem is considered. Many different approaches to the link prediction have been proposed for the past decade, but with the objective of showing the potential merit of the statistically significance attribute associations, we simply use the Jaccard coefficient proposed in~\cite{liben2007link} and compare the effects of using statistically significant attribute associations and frequent ones. Given a pair of nodes without an edge, we compute the prediction score by combining the Jaccard coefficient $J(u, v)$ and the score $S(u, v)$ resulted from either the significance or the normalized frequency of an attribute association between the nodes as follows
\begin{eqnarray}
	pred(u, v) = \tau \cdot J(u, v) + (1 - \tau) \cdot S(u, v)
\end{eqnarray}
and if it is over a given threshold then we predict that $u$ and $v$ will form a new link. We take two snapshots of the \textit{DBLP co-authorship network} (Mar 2015 and Mar 2016) and all the newly created links between the two snapshots are used for the positive samples. Similarly, a set of pairs of nodes that do not have an edge in both the snapshots are used for the negative samples. Since the number of negative samples far outweighs the number of positive samples, we do negative subsampling with the ratio of $1:5$ (five negatives per one positive). In \figurename~\ref{fig:link_prediction}, we report the ROC curves for two different methods, \texttt{Jaccard+Significant}, and \texttt{Jaccard+Frequent}. As shown in \figurename~\ref{fig:link_prediction}, the link prediction can more benefit from employing the attribute information and the statistically significant attribute associations can achieve higher performance rather than the frequent ones.

\section{Conclusion}
\label{sec:conclusion}

We defined a problem of mining statistically significant attribute associations using \textit{Association Graph}, which keeps the locality of attribute associations and carries the significant relationships between the sets of attribute values. And we proposed a novel, two-step iterative algorithm that efficiently and effectively generates an \textit{Association Graph} from the original graph. The experiments are conducted on two real world datasets, and we ran some qualitative analysis on the results, confirming that our algorithm effectively finds the significant associations, which cannot be uncovered by conventional frequent association mining. Also we ran extensive scalability experiments on synthetic datasets, and confirmed that the algorithm is of polynomial running time in the number of nodes. Lastly, applying the results from one of the real world datasets to the link prediction task, and we showed how the statistically significant attribute associations can be used in practice.

For future work, we plan to investigate how we can exploit resulting \textit{Association Graph} better, e.g., its locality preserving property, and if we can come up with a linear time algorithm or a distributed algorithm, which can be run on large-scale graphs.

\bibliographystyle{IEEEtran}
\bibliography{IEEEabrv,reference}
%
%
%

\end{document}